\documentclass[12pt]{article}
\usepackage{graphicx}
\usepackage[cp1251]{inputenc}

\begin{document}
 \noindent {\footnotesize\it
   Astronomy Letters, 2020, Vol. 46, No 4, pp. 245--255.}
 \newcommand{\dif}{\textrm{d}}

 \noindent
 \begin{tabular}{llllllllllllllllllllllllllllllllllllllllllllll}
 & & & & & & & & & & & & & & & & & & & & & & & & & & & & & & & & & & & & & \\\hline\hline
 \end{tabular}

  \vskip 0.5cm
  \centerline{\bf\large Analysis of Close Stellar Encounters with the Solar System}
   \bigskip
   \bigskip
  \centerline
 {V.V. Bobylev\footnote [1]{e-mail: vbobylev@gaoran.ru}  and A. T. Bajkova}
   \bigskip

  \centerline{\small\it Pulkovo Astronomical Observatory, Russian Academy of Sciences,}

  \centerline{\small\it Pulkovskoe sh. 65, St. Petersburg, 196140 Russia}
 \bigskip
 \bigskip
 \bigskip

 {
{\bf Abstract}---We consider 36 candidates for close (within 1 pc) encounters with the Solar system. These
stars have been selected in accordance with the results of an analysis of their motion obtained by various
authors. For most stars from this list the kinematic characteristics have been taken from the Gaia DR2
catalogue. The parameters of the encounters of these stars with the Solar system have been calculated
using three methods: (1) the linear one, (2) the epicyclic one, and (3) by integrating the orbits in an
axisymmetric potential. We have concluded that the epicyclic method works quite well only on a time
interval no longer than $\pm$1 Myr. Based on the third method, in good agreement with the first method,
for the first time we have obtained the following estimates of the encounter parameters for the star
Gaia DR2 3130033734235815424: $t_{min}=-0.62\pm0.12$ Myr and $d_{min}=0.30\pm0.10$ pc.
  }

 % DOI: 10.1134/S1063773720040039

 \subsection*{INTRODUCTION}
The interest in close (within 1--2 pc) encounters of field stars with the Solar system stems from the
fact that the so-called comet shower from the outer boundaries of the Oort cloud (Oort 1950) toward
the major planets and, in particular, the Earth can emerge. As simulations show (Dybczy\'nski 2002,
2005; Martinez-Barbosa et al. 2017), apart from stellar flybys, the Oort comet cloud is subject to perturbations from giant molecular clouds and experiences an effect from the Galaxy’s attraction.

A practical search for close stellar encounters with the Solar system was carried out, for example, by
Revina (1988), Matthews (1994), and M\"ull\"ari and Orlov (1996). The then used ground-based catalogues
were not distinguished by a high accuracy of stellar parallaxes and proper motions. Nevertheless,
thanks to these authors, some candidates were revealed, which were subsequently confirmed by more
accurate data. These include, for example, Proxima Centauri, the $\alpha$ Centauri system, or the star GJ\,905.

Based on data from the Hipparcos (1997) catalogue, this problem was solved by Garcia-S\'anchez
et al. (1999, 2001), Bobylev (2010a, 2010b), Anderson and Francis (2012), Dybczy\'nski and Berski (2015),
Bailer-Jones (2015), and Feng and Bailer-Jones (2015).

Based on data from the Gaia catalogue (Prusti et al. 2016), its first release --- Gaia TGAS (Tycho-Gaia Astrometric Solution, Lindegren et al. 2016), the search for close encounters was conducted by
Berski and Dybczy\'nski (2016), Bobylev and Bajkova (2017), and R. de la Fuente Marcos and C. de la Fuente Marcos (2018). Several candidates for a very close flyby, namely for an injection into the Oort cloud (to distances less than 0.5 pc), were detected as a result of these efforts. The record-holder is the star GJ 710 (Garcia-S\'anchez et al. 2001; Bobylev 2010a; Berski and Dybczy\'nski 2016; Bailer-Jones 2018). Another
example is the low-mass (M9.5+T5) binary system WISE J072003.20--084651.2 detected by Mamajek
et al. (2015).

Finally, based on even more accurate Gaia DR2 data (Brown et al. 2018; Lindegren et al. 2018), such an analysis was performed, for example, by Bailer-Jones et al. (2018), Darma et al. (2019), Torres et al. (2019), and Wysocza\'nska et al. (2020). As a result, we have: $\sim$3000 candidates, $\sim$30 stars, and 5--6 stars can have encounters with the Solar system within 5, 1, and 0.25 pc, respectively, on a time interval of $\pm$5 Myr. In particular, Wysocza\'nska et al. (2020) found the star ALS 9243 that could approach the solar orbit to a distance of 0.25 pc 2.5 Myr ago, while the record-holder star GJ 710, according to the estimate by Bailer-Jones et al. (2018), may approach the Solar system to a distance of $\sim$0.05 pc in 1.2 Myr.

Stellar trajectories are constructed by various methods. The following ones are commonly applied:
(i) the linear method, (ii) the epicyclic approximation, or (iii) using an appropriate model Galactic gravitational potential. Large (more than 1 pc) discrepancies between the results of an analysis of the encounter parameters for ALS 9243 (Wysocza\'nska et al. 2020) and several more stars from Darma et al. (2019) that we got by applying various approaches served as a stimulus for writing this paper.

The goal of our paper is to use these three methods to analyze $\sim$30 stars that, according to various authors, have encounters with the Solar system within 1 pc. In our opinion, such a work is topical in the runup
to the release of the Gaia DR3 catalogue.

 \subsection*{ORBIT CONSTRUCTION METHODS}
In a rectangular coordinate system with the center in the Sun the $x$ axis is directed toward the Galactic
center, the $y$ axis is in the direction of Galactic rotation, and the $z$ axis is directed to the north Galactic pole. Then, $x=r\cos l\cos b, y=r\sin l\cos b,$ and $z=r\sin b,$ where $r = 1/\pi$ is the stellar heliocentric distance in kpc that we calculate via the stellar parallax $\pi$ in mas. Note that in this paper we use stars with relative parallax errors less than 10\% and, therefore, there is no need to take into account the Lutz–Kelker bias (Lutz and Kelker 1973).

The line-of-sight velocity $V_r$ and the two tangential velocity components $V_l=4.74r\mu_l\cos b$ and $V_b=4.74r\mu_b$ along the Galactic longitude $l$ and latitude $b,$ respectively, expressed in km s$^{-1}$ are known from observations. Here, the coefficient 4.74 is the ratio of the number of kilometers in an astronomical unit to the number of seconds in a tropical year. The proper motion components $\mu_l\cos b$ and $\mu_b$ are expressed in mas yr$^{-1}$.

The velocities $U, V,$ and $W,$ where $U$ is directed away from the Sun toward the Galactic center, $V$
is in the direction of Galactic rotation, and $W$ is directed to the north Galactic pole, are calculated via
the components $V_r, V_l,$ and $V_b,$ respectively:
 \begin{equation}
 \begin{array}{lll}
 U=V_r\cos l\cos b-V_l\sin l-V_b\cos l\sin b,\\
 V=V_r\sin l\cos b+V_l\cos l-V_b\sin l\sin b,\\
 W=V_r\sin b                +V_b\cos b.
 \label{UVW}
 \end{array}
 \end{equation}
We use three methods to analyze the close encounters of field stars with the Solar system. The first method is
based on the linear approximation, the second method consists in analyzing the epicyclic orbits of stars, and
in the third method the stellar and solar orbits are constructed using an axisymmetric model Galactic
gravitational potential.

 \subsubsection*{Linear Method}
According to Matthews (1994), the minimum distance between the stellar and solar trajectories $d_{\rm min}$
at the encounter time $t_{\rm min}$ can be found from the following relations:
 \begin{equation}
 \renewcommand{\arraystretch}{1.4}
 \begin{array}{lll}
   d_{\rm min}=r /\sqrt{1+(V_r/V_t)^2},\\
   t_{\rm min}=r V_r/(V^2_t+V^2_r),
 \label{lin}
 \end{array}
 \end{equation}
where $V_t=\sqrt{V^2_l+V^2_b}$ is the stellar velocity perpendicular to the line of sight.

 \subsubsection*{Epicyclic Approximation}
The epicyclic approximation (Lindblad 1927) allows the orbits of stars to be constructed in a coordinate
system rotating around the Galactic center. We apply the method in the form given by Fuchs
et al. (2006):
 \begin{equation}
 \renewcommand{\arraystretch}{1.8}
 \begin{array}{lll}\displaystyle
 x(t)= x_0+{U_0\over \displaystyle \kappa}\sin(\kappa t)   
      +{\displaystyle V_0\over \displaystyle 2B}(1-\cos(\kappa t)),  \\
 y(t)= y_0+2A \biggl(x_0+{\displaystyle V_0\over\displaystyle 2B}\biggr) t
       -{\displaystyle \Omega_0\over \displaystyle B\kappa} V_0\sin(\kappa t)
       +{\displaystyle 2\Omega_0\over \displaystyle \kappa^2} U_0(1-\cos(\kappa t)),\\
 z(t)= {\displaystyle W_0\over \displaystyle \nu} \sin(\nu t)+z_0\cos(\nu t),
 \label{EQ-Epiciclic}
 \end{array}
 \end{equation}
where $t$ is the time in Myr (we proceed from pc/Myr=0.978 km s$^{-1}$); $A$ and $B$ are the Oort constants;
$\kappa=\sqrt{-4\Omega_0 B}$ is the epicyclic frequency; $\Omega_0$ is the angular velocity of the Galactic rotation of the local standard of rest, $\Omega_0=A-B$; $\nu=\sqrt{4\pi G \rho_0}$ is the vertical oscillation frequency, where $G$ is the gravitational constant and $\rho_0$ is the stellar density in the solar neighborhood.

The parameters $x_0,y_0,z_0$ and $U_0,V_0,W_0$ in the system of equations (3) denote the current stellar
positions and velocities. The Sun’s height above the Galactic plane is taken to be $h_\odot=16$ pc (Bobylev
and Bajkova 2016a). We calculate the velocities $U, V,$ and $W$ relative to the local standard of rest
using $(U_\odot,V_\odot,W_\odot)=(11.1,12.2,7.3)$ km s$^{-1}$ from Sch\"onrich et al. (2010).

We adopted $\rho_0=0.1~M_\odot/$pc$^{-3}$ (Holmberg and Flinn 2004), which gives $\nu=74$ km s$^{-1}$ kpc$^{-1}$. The following Oort constants close to their present-day estimates are used: $A=18.5$ km s$^{-1}$ kpc$^{-1}$ and $B=-11.0$ km s$^{-1}$ kpc$^{-1}$ (Rastorguev et al. 2017). Note that we neglect the star–Sun gravitational interaction.

For each star we calculate the encounter parameter between the stellar and solar orbits $d(t)=\sqrt{\Delta x^2(t)+\Delta y^2(t)+\Delta z^2(t)}$. Next, we determine $d_{\rm min}$ at the encounter time $t_{\rm
min}$ from these data.

 \subsubsection*{Model Gravitational Potential}
The axisymmetric Galactic potential is represented as a sum of three components---a central
spherical bulge $\Phi_b(r(R,Z))$, a disk $\Phi_d(r(R,Z))$, and a massive spherical dark matter halo $\Phi_h(r(R,Z))$:
 \begin{equation}
 \begin{array}{lll}
  \Phi(R,Z)=\Phi_b(r(R,Z))+\Phi_d(r(R,Z))+\Phi_h(r(R,Z)).
 \label{pot}
 \end{array}
 \end{equation}
Here, we use a cylindrical coordinate system $(R,\psi,Z)$ with the coordinate origin at the Galactic center. In a rectangular coordinate system $(X,Y,Z)$ the distance to a star (spherical radius) will be $r^2=X^2+Y^2+Z^2=R^2+Z^2$, with the $X$ axis being directed away from the Sun toward the Galactic center, the $Y$ axis being perpendicular to the $X$ axis in the direction of Galactic rotation, and the $Z$ axis being perpendicular to the Galactic $XY$ plane and directed toward the north Galactic pole. The gravitational potential is expressed in units of 100 km$^2$ s$^{-2},$ the distances are in kpc, and the masses are in units of the Galactic mass $M_{\rm gal}=2.325\times 10^7 M_\odot$ corresponding to the gravitational constant $G=1.$

The bulge, $\Phi_b(r(R,Z)),$ and disk, $\Phi_d(r(R,Z)),$ potentials are represented in the form proposed by
Miyamoto and Nagai (1975):
 \begin{equation}
  \Phi_b(r)=-\frac{M_b}{(r^2+b_b^2)^{1/2}},
  \label{bulge}
 \end{equation}
 \begin{equation}
 \Phi_d(R,Z)=-\frac{M_d}{\Biggl[R^2+\Bigl(a_d+\sqrt{Z^2+b_d^2}\Bigr)^2\Biggr]^{1/2}},
 \label{disk}
\end{equation}
where $M_b$ and $M_d$ are the masses of the components, $b_b, a_d,$ and $b_d$ are the scale lengths of the components in kpc. The halo component is represented according to Navarro et al. (1997):
 \begin{equation}
  \Phi_h(r)=-\frac{M_h}{r}\ln {\Biggl(1+\frac{r}{a_h}\Biggr)}.
 \label{halo-III}
 \end{equation}
The parameters of the model Galactic potential (5)--(7) are given in Table 1. In Bajkova and Bobylev
(2016b) the model (5)--(7) is designated as model III. The total mass of the Galaxy within 200 kpc in this
model is $M_{\rm 200}=(0.75\pm0.19)\times10^{12}M_\odot.$

%%%%%%%%%%%%%%%%%%%%%%%%%%%%%%%%%%%%%%%%%%%%%%%%%%%%%%%%
 {\begin{table}[t]                                    %% t1.
 \caption[]
 {\small\baselineskip=1.0ex
Parameters of the model Galactic potential from Bajkova and Bobylev (2016b), 
 $M_{\rm gal}=2.325\times 10^7 M_\odot$
  }
 \label{t:model-III}
 \begin{center}\begin{tabular}{|c|c|r|}\hline
 Parameters           &  Model III\\\hline
 $M_b$($M_{\rm gal}$) &    443$\pm27$  \\
 $M_d$($M_{\rm gal}$) &   2798$\pm84$ \\
 $M_h$($M_{\rm gal}$) &  12474$\pm3289$ \\
 $b_b$(kpc)       & 0.2672$\pm0.0090$ \\
 $a_d$(kpc)       &   4.40$\pm0.73$ \\
 $b_d$(kpc)       & 0.3084$\pm0.0050$ \\
 $a_h$(kpc)       &    7.7$\pm2.1$ \\\hline
 \end{tabular}\end{center}\end{table}}
%%%%%%%%%%%%%%%%%%%%%%%%%%%%%%%%%%%%%%%%%%%%%%%%%%%%%%%%

The equations of motion for a test particle in a Galactic potential appear as follows:
\begin{equation}
 \begin{array}{llllll}
 \dot{X}=p_X,\quad
 \dot{Y}=p_Y,\quad
 \dot{Z}=p_Z,\\
 \dot{p}_X=-\partial\Phi/\partial X,\\
 \dot{p}_Y=-\partial\Phi/\partial Y,\\
 \dot{p}_Z=-\partial\Phi/\partial Z,
 \label{eq-motion}
 \end{array}
\end{equation}
where$p_X, p_Y, p_Z$ are the canonical momenta, the dot denotes a time derivative. The fourth-order
Runge--Kutta algorithm was used to integrate Eqs. (8).

In the rectangular Galactic coordinate system the initial test particle positions and velocities are determined
from the formulas
\begin{equation}
 \begin{array}{llllll}
 X=R_0-x_0, Y=y_0, Z=z_0+h_\odot,\\
 U=-(U_0+U_\odot),\\
 V=V_0+V_\odot+V_{circ},\\
 W=W_0+W_\odot,
 \label{init}
 \end{array}
\end{equation}
where $(x_0,y_0,z_0,U_0,V_0,W_0)$ are the initial test particle positions and space velocities in the heliocentric coordinate system and the circular rotation velocity of the solar neighborhood in our potential is $V_{\rm circ}=244$ km s$^{-1}.$

As above, for each star we calculate the encounter parameter between the stellar and solar orbits $d(t)=\sqrt{\Delta X^2(t)+\Delta Y^2(t)+\Delta Z^2(t)}$. Then, we determine $d_{\rm min}$ at the encounter time $t_{\rm min}$.

We estimate the errors in $d_{\rm min}$ and $t_{\rm min}$ by the Monte Carlo method. Here, the errors in the stellar parameters are assumed to be distributed normally with a dispersion $\sigma.$ The errors are added to the equatorial coordinates, proper motion components, parallaxes, and line-of-sight velocities of the stars.

%%%%%%%%%%%%%%%%%%%%%%%%%%%%%%%%%%%%%%%%%%%%%%%%%
 \begin{table}[p]                               %T~2
 \caption[]{\small\baselineskip=1.0ex\protect
 Input data on the stars }
 \begin{center}
 %\begin{tabular}{|r|c|c|c|c|c|c|c|}\hline
 \begin{tabular}{|r|r|r|r|r|r|r|r|}\hline
 \label{tab-data}

 \def\baselinestretch{1}\normalsize\small

 Gaia DR2/alternative &  $\pi,$        & $\mu_\alpha\cos\delta,$ & $\mu_\delta,$ & $V_r,$ \\
                      &    mas         &   mas yr$^{-1}$ &   mas yr$^{-1}$ &  km s$^{-1}$  \\\hline
              GJ 710 & $ 52.52\pm0.05$ & $ -0.46\pm0.08$ & $ -0.03\pm0.07$ & $-14.5 \pm0.0$ \\
  955098506408767360 & $ 34.51\pm0.61$ & $  0.11\pm0.92$ & $  0.82\pm0.79$ & $ 38.5 \pm2.1$ \\
 5700273723303646464 & $ 15.67\pm1.08$ & $  0.16\pm1.39$ & $ -0.21\pm1.36$ & $ 38.0 \pm0.9$ \\
            ALS 9243 & $ 10.56\pm0.40$ & $ -0.11\pm0.58$ & $ -0.10\pm0.63$ & $  40  \pm8 $  \\
 2946037094755244800 & $ 25.63\pm1.12$ & $ -0.33\pm1.47$ & $ -1.30\pm1.21$ & $ 42.1 \pm3.2$ \\
 5571232118090082816 & $ 10.20\pm0.02$ & $  0.10\pm0.04$ & $  0.41\pm0.04$ & $ 82.18\pm0.47$ \\
         WISE J07200 & $147.1 \pm1.2 $ & $ -46.0\pm4.0 $ & $-116.5\pm2.2 $ & $  82.4\pm0.3 $ \\
  154460050601558656 & $ 11.26\pm0.67$ & $ -2.08\pm0.81$ & $ -0.55\pm0.44$ & $  -233\pm 9 $ \\
 4071528700531704704 & $ 50.40\pm0.89$ & $ -0.71\pm1.45$ & $ -8.88\pm1.25$ & $   -45\pm17$  \\
 4472507190884080000 & $ 10.34\pm0.61$ & $  0.10\pm0.86$ & $  0.50\pm0.63$ & $   -52\pm15$  \\
 3376241909848155520 & $ 27.15\pm1.09$ & $ -2.04\pm1.89$ & $  5.60\pm1.75$ & $  79.9\pm5.6$ \\
 1791617849154434688 & $ 11.46\pm0.04$ & $ -0.38\pm0.06$ & $ -0.79\pm0.07$ & $ 56.29\pm.48$ \\
  510911618569239040 & $ 13.20\pm0.04$ & $  0.56\pm0.04$ & $  0.01\pm0.05$ & $ 26.45\pm.35$ \\
 4265426029901799552 & $ 32.02\pm0.88$ & $ -5.30\pm1.41$ & $ -2.65\pm1.27$ & $  6.58\pm.19$ \\
 4252068750338781824 & $ 38.84\pm0.61$ & $ -4.47\pm0.94$ & $ -3.60\pm0.79$ & $    28\pm14$  \\
 5261593808165974784 & $ 15.29\pm0.02$ & $ -0.20\pm0.03$ & $ -2.32\pm0.05$ & $ 71.05\pm0.88$ \\
 1949388868571283200 & $  3.93\pm0.04$ & $ -0.33\pm0.04$ & $ -0.73\pm0.06$ & $   347\pm7$   \\
 3105694081553243008 & $ 35.69\pm0.97$ & $  6.16\pm1.60$ & $  4.19\pm1.39$ & $  38.4\pm1.9$ \\
 3996137902634436480 & $ 39.68\pm1.07$ & $  2.41\pm1.96$ & $-10.23\pm2.10$ & $ -38.4\pm2.3$ \\
 3260079227925564160 & $ 32.16\pm0.06$ & $ -3.41\pm0.10$ & $ -4.94\pm0.06$ & $ -33.4\pm0.4$  \\
 5231593594752514304 & $ 15.32\pm0.03$ & $-29.87\pm0.06$ & $ -0.01\pm0.05$ & $  -716\pm1 $  \\
 3458393840965496960 & $ 13.17\pm1.05$ & $  1.50\pm1.47$ & $ -2.36\pm1.29$ & $    87\pm20$  \\
         Proxima Cen & $771.64\pm2.60$ & $-3775.7\pm1.6$ & $ 765.5\pm2.0 $ & $ -25.1\pm0.9 $ \\
     $\alpha$~Cen AB & $754.81\pm4.11$ & $-3643.0\pm2.0$ & $ 697.0\pm2.0 $ & $ -24.7\pm0.4 $ \\
 3972130276695660288 & $ 59.94\pm0.05$ & $-21.03\pm0.11$ & $  6.52\pm0.10$ & $ 31.80\pm0.73$ \\
              GJ 905 & $316.95\pm0.12$ & $112.69\pm0.15$ & $-1592.1\pm0.1$ & $ -78.0\pm0.4 $ \\
 2926732831673735168 & $  8.72\pm0.04$ & $ -0.81\pm0.06$ & $  0.57\pm0.06$ & $ 66.49\pm0.25$ \\
 6724929671747826816 & $ 17.04\pm0.49$ & $  2.89\pm0.76$ & $  1.20\pm0.66$ & $ -54.8\pm1.1$ \\
          AC+79 3888 & $190.26\pm0.05$ & $748.11\pm0.10$ & $480.60\pm0.08$ & $-111.6\pm0.2$ \\
  939821616976287104 & $ 19.02\pm0.07$ & $-45.67\pm0.10$ & $ -1.91\pm0.10$ & $ 568.3\pm0.8$ \\
 2924378502398307840 & $  6.07\pm0.03$ & $  0.74\pm0.03$ & $  0.11\pm0.05$ & $  86.9\pm1.0$ \\
 6608946489396474752 & $  7.87\pm0.05$ & $ -0.62\pm0.07$ & $ -0.25\pm0.08$ & $  44.2\pm0.57$ \\
   52952724810126208 & $ 47.86\pm1.82$ & $  0.04\pm1.50$ & $ -4.60\pm1.48$ & $  37.8\pm3.4$  \\
 3130033734235815424 & $ 38.37\pm0.90$ & $  3.25\pm1.43$ & $ -2.34\pm1.25$ & $    42\pm7$   \\
  969867803725057920 & $ 50.74\pm1.30$ & $  8.93\pm1.95$ & $  9.21\pm1.80$ & $    41\pm4$   \\
  365942724131566208 & $ 56.29\pm1.96$ & $-10.06\pm1.97$ & $ 12.41\pm2.00$ & $   -29\pm5$   \\
 \hline
 \end{tabular} \end{center} \end{table}
%%%%%%%%%%%%%%
%%%%%%%%%%%%%%%%%%%%%%%%%%%%%%%%%%%%%%%%%%%%%%%%%
 \begin{table}[p]                               %T~3
 \caption[]{\small\baselineskip=1.0ex\protect
 Parameters of the stellar encounters with the Solar system }
 \begin{center}
 \begin{tabular}{|r|rr|rr|rr|rr|}\hline
 \label{tab-rez}
    Gaia DR2/alternative & $t_{\rm min},$ & $d_{\rm min},$ & $t_{\rm min},$ & $d_{\rm min},$ & $t_{\rm min},$ & $d_{\rm min},$ & $\sigma_t,$ & $\sigma_d,$\\
                     &  Myr   &  pc        &   Myr   &  pc      &  Myr     &  pc     &  Myr  &  pc  \\\hline
                      &     (1)&            &      (2)&          &       (3)&         &       &      \\\hline
              GJ 710 & $ 1.316$ & $ .055 $ & $ 1.344$ & $ .052$ & $ 1.320$ & $.016$ & .040  & .009 \\
  955098506408767360 & $ -.752$ & $ .085 $ & $ -.754$ & $ .105$ & $ -.755$ & $.070$ & .046  & .064 \\
 5700273723303646464 & $-1.677$ & $ .134 $ & $-1.633$ & $2.052$ & $-1.678$ & $.195$ & .168  & .938 \\
            ALS 9243 & $-2.367$ & $ .154 $ & $-2.272$ & $1.218$ & $-2.370$ & $.211$ & .521  & .798 \\
 2946037094755244800 & $ -.927$ & $ .230 $ & $ -.911$ & $ .286$ & $ -.928$ & $.227$ & .091  & .234 \\
 5571232118090082816 & $-1.193$ & $ .231 $ & $-1.177$ & $2.856$ & $-1.193$ & $.171$ & .007  & .030 \\
         WISE J07200 & $ -.082$ & $ .333 $ & $ -.082$ & $ .328$ & $ -.083$ & $.370$ & .001  & .012 \\
  154460050601558656 & $  .381$ & $ .345 $ & $  .380$ & $ .351$ & $  .382$ & $.329$ & .032  & .146 \\
 4071528700531704704 & $  .446$ & $ .374 $ & $  .447$ & $ .316$ & $  .447$ & $.377$ & .151  & .135 \\
 4472507190884080000 & $ 1.852$ & $ .434 $ & $ 1.905$ & $ .806$ & $ 1.854$ & $.331$ & .470  & .625 \\
 3376241909848155520 & $ -.461$ & $ .479 $ & $ -.459$ & $ .373$ & $ -.462$ & $.467$ & .043  & .156 \\
 1791617849154434688 & $-1.549$ & $ .562 $ & $-1.521$ & $3.692$ & $-1.549$ & $.599$ & .015  & .073 \\
  510911618569239040 & $-2.863$ & $ .576 $ & $-3.009$ & $3.391$ & $-2.863$ & $.420$ & .041  & .087 \\
 4265426029901799552 & $ -.670$ & $ .588 $ & $ -.663$ & $ .638$ & $-.672$ & $.596$ & .025  & .172 \\
 4252068750338781824 & $  .930$ & $ .652 $ & $  .943$ & $ .671$ & $ .932$ & $.652$ & .532  & .391 \\
 5261593808165974784 & $ -.920$ & $ .664 $ & $ -.927$ & $ .874$ & $-.920$ & $.650$ & .011  & .019 \\
 1949388868571283200 & $ -.732$ & $ .709 $ & $ -.730$ & $5.948$ & $-.733$ & $.699$ & .017  & .235 \\
 3105694081553243008 & $ -.730$ & $ .722 $ & $ -.720$ & $ .740$ & $-.731$ & $.711$ & .046  & .126 \\
 3996137902634436480 & $  .655$ & $ .822 $ & $  .656$ & $ .808$ & $ .656$ & $.801$ & .037  & .179 \\
 3260079227925564160 & $  .931$ & $ .823 $ & $  .934$ & $ .803$ & $ .931$ & $.817$ & .010  & .014 \\
 5231593594752514304 & $  .091$ & $ .842 $ & $  .091$ & $ .976$ & $ .092$ & $.813$ & .002  & .017 \\
 3458393840965496960 & $ -.876$ & $ .882 $ & $ -.881$ & $1.333$ & $-.878$ & $.824$ & .291  & .544 \\
         Proxima Cen & $  .027$ & $ .889 $ & $  .027$ & $ .889$ & $ .028$ & $.877$ & .015  & .001 \\
     $\alpha$~Cen AB & $  .028$ & $ .909 $ & $  .028$ & $ .909$ & $ .028$ & $.889$ & .000  & .011 \\
 3972130276695660288 & $ -.523$ & $ .912 $ & $ -.522$ & $ .981$ & $-.523$ & $.898$ & .011  & .022 \\
              GJ 905 & $  .037$ & $ .923 $ & $  .037$ & $ .924$ & $ .037$ & $.908$ & .000  & .004 \\
 2926732831673735168 & $-1.725$ & $ .925 $ & $-1.674$ & $3.431$ & $-1.725$ & $.970$ & .013  & .075 \\
 6724929671747826816 & $ 1.071$ & $ .932 $ & $ 1.066$ & $1.295$ & $ 1.073$ & $.869$ & .043  & .173 \\
          AC+79 3888 & $  .045$ & $1.023 $ & $  .045$ & $1.025$ & $  .046$ & $1.003$ & .000  & .002 \\
  939821616976287104 & $ -.092$ & $1.054 $ & $ -.093$ & $1.067$ & $ -.093$ & $1.108$ & .001  & .019 \\
 2924378502398307840 & $-1.893$ & $1.098 $ & $-1.833$ & $4.629$ & $-1.892$ & $.917$ & .023  & .103 \\
 6608946489396474752 & $-2.875$ & $1.161 $ & $-2.818$ & $2.152$ & $-2.845$ & $.456$ & .042  & .154 \\
   52952724810126208 & $ -.553$ & $ .252 $ & $ -.554$ & $ .323$ & $-.553$ & $.205$ & .060  & .086 \\
 3130033734235815424 & $ -.618$ & $ .306 $ & $ -.613$ & $ .301$ & $-.620$ & $.299$ & .116  & .096 \\
  969867803725057920 & $ -.482$ & $ .578 $ & $ -.484$ & $ .590$ & $-.484$ & $.576$ & .057  & .102 \\
  365942724131566208 & $  .619$ & $ .834 $ & $  .615$ & $ .913$ & $ .619$ & $.829$ & .089  & .182 \\
 \hline
 \end{tabular} \end{center}
 {\small (1) the linear method, (2) the epicyclic method, and (3) the axisymmetric potential.}
 \end{table}
%%%%%%%%%%%%%%
%%%%%%%%%%%%%%%%%%%%%%%%%%%%%%%%%%%%%%%%%%%%%%%%%
 \begin{table}[p]                               %T~4
 \caption[]{\small\baselineskip=1.0ex\protect
 ``Method 1 minus Bailer-Jones'' and ``method 3 minus Bailer-Jones'' (c, d) parameter differences
 }
 \begin{center}
 \begin{tabular}{|r|rr|rr|rr}\hline
 \label{tab-BJones}
    Gaia DR2/alternative & $\Delta t_{\rm min},$ Myr & ~~$\Delta d_{\rm min},$ pc & $\Delta t_{\rm min},$ Myr & ~~$\Delta d_{\rm min},$ pc\\\hline
                     &  (1)--BJ&         &  (3)--BJ&         \\\hline
              GJ 710 & $ .095$ & $-.029$ & $ .100$ & $-.036$ \\
  955098506408767360 & $ .064$ & $ .042$ & $ .060$ & $-.027$ \\
 5700273723303646464 & $ .169$ & $-.093$ & $ .168$ & $-.032$ \\
 2946037094755244800 & $ .127$ & $ .129$ & $ .116$ & $-.126$ \\
 5571232118090082816 & $-.020$ & $ .132$ & $-.020$ & $-.026$ \\
  154460050601558656 & $ .049$ & $ .120$ & $ .050$ & $-.104$ \\
 4071528700531704704 & $ .178$ & $ .160$ & $ .179$ & $ .164$ \\
 4472507190884080000 & $ .642$ & $ .220$ & $ .644$ & $ .117$ \\%==
 3376241909848155520 & $ .055$ & $ .205$ & $ .054$ & $ .141$ \\
 1791617849154434688 & $-.016$ & $ .059$ & $-.016$ & $ .096$ \\
  510911618569239040 & $-.010$ & $ .208$ & $-.010$ & $ .052$ \\
 4265426029901799552 & $ .016$ & $ .220$ & $ .014$ & $ .228$ \\
 4252068750338781824 & $ .478$ & $ .300$ & $ .470$ & $ .300$ \\%==
 5261593808165974784 & $-.004$ & $ .056$ & $-.004$ & $ .042$ \\
 1949388868571283200 & $ .006$ & $ .181$ & $ .005$ & $ .108$ \\
 3105694081553243008 & $ .060$ & $ .216$ & $ .059$ & $ .195$ \\
 3996137902634436480 & $ .011$ & $ .275$ & $ .012$ & $ .254$ \\
 3260079227925564160 & $ .041$ & $ .025$ & $ .041$ & $ .019$ \\
 5231593594752514304 & $-.002$ & $ .035$ & $-.003$ & $ .006$ \\
 3458393840965496960 & $ .543$ & $ .476$ & $ .541$ & $ .428$ \\%==
 3972130276695660288 & $ .007$ & $ .057$ & $ .007$ & $ .037$ \\
 2926732831673735168 & $-.027$ & $ .105$ & $-.027$ & $ .150$ \\
 6724929671747826816 & $ .085$ & $ .420$ & $ .087$ & $ .357$ \\%==
  939821616976287104 & $-.001$ & $ .080$ & $-.002$ & $ .119$ \\
 2924378502398307840 & $-.013$ & $ .294$ & $-.012$ & $ .114$ \\
 6608946489396474752 & $-.049$ & $ .860$ & $-.021$ & $ .155$ \\
 \hline
 \end{tabular} \end{center}
 \end{table}
%%%%%%%%%%%%%%

 \subsection*{DATA}
Such input data on our 36 stars as the trigonometric parallaxes, proper motion components, and line of-sight velocities are given in Table 2. The sample was produced as follows.

(1) From Table 2 in Bailer-Jones et al. (2018) we took 26 stars that approach the Solar system within
1 pc on the time interval from $-$3 to +3 Myr.

(2) We added the star ALS 9243 to the list of candidates based on the analysis by Wysocza\'nska et al. (2020).

(3) The low-mass binary star WISE J072003.20--084651.2AB (Scholz's star) was first revealed as an
interesting candidate for close encounters by Mamajek et al. (2015). It is absent in the Gaia DR2
catalogue. We took a new estimate of its dynamical parallax and absolute proper motion from Dupuy
et al. (2019).

(4) We added the data on the four well-known stars Proxima Cen, $\alpha$~Cen AB, GJ 905, and
AC+79 3888 based on the previous papers of various authors (Matthews 1994; M\"ull\"ari and Orlov 1996;
Garcia-S\'anchez et al. 1999; Bobylev 2010a, 2010b).

(5)We added four more stars,
Gaia DR2 52952724810126208,
Gaia DR2 3130033734235815424,
Gaia DR2 969867803725057920, and
Gaia DR2 365942724131566208,
to our list based on data from Darma et al. (2019), where the following indices are given for them:
ID 298, ID 291, ID 299, and ID 300. Here, we ran into the absence of star numbers from the Gaia DR2
catalogue in Darma et al. (2019). Although the list of candidates consists of 11 stars, we took only those
that had been identified with the Gaia DR2 catalogue by their coordinates. The stars
Gaia DR2 3130033734235815424,
Gaia DR2 969867803725057920, and
Gaia DR2 365942724131566208
are of great interest in that the line-of-sight velocities from the LAMOST program are given for them in
Darma et al. (2019) for the first time.

%%%%%%%%%%%%%%%%%%%%%%%%%%%%%%%%%%%%%%%%%%%%%%%%%%%%%%%%%% Fig.1
 \begin{figure} [p] {\begin{center}
 \includegraphics[width=140mm]{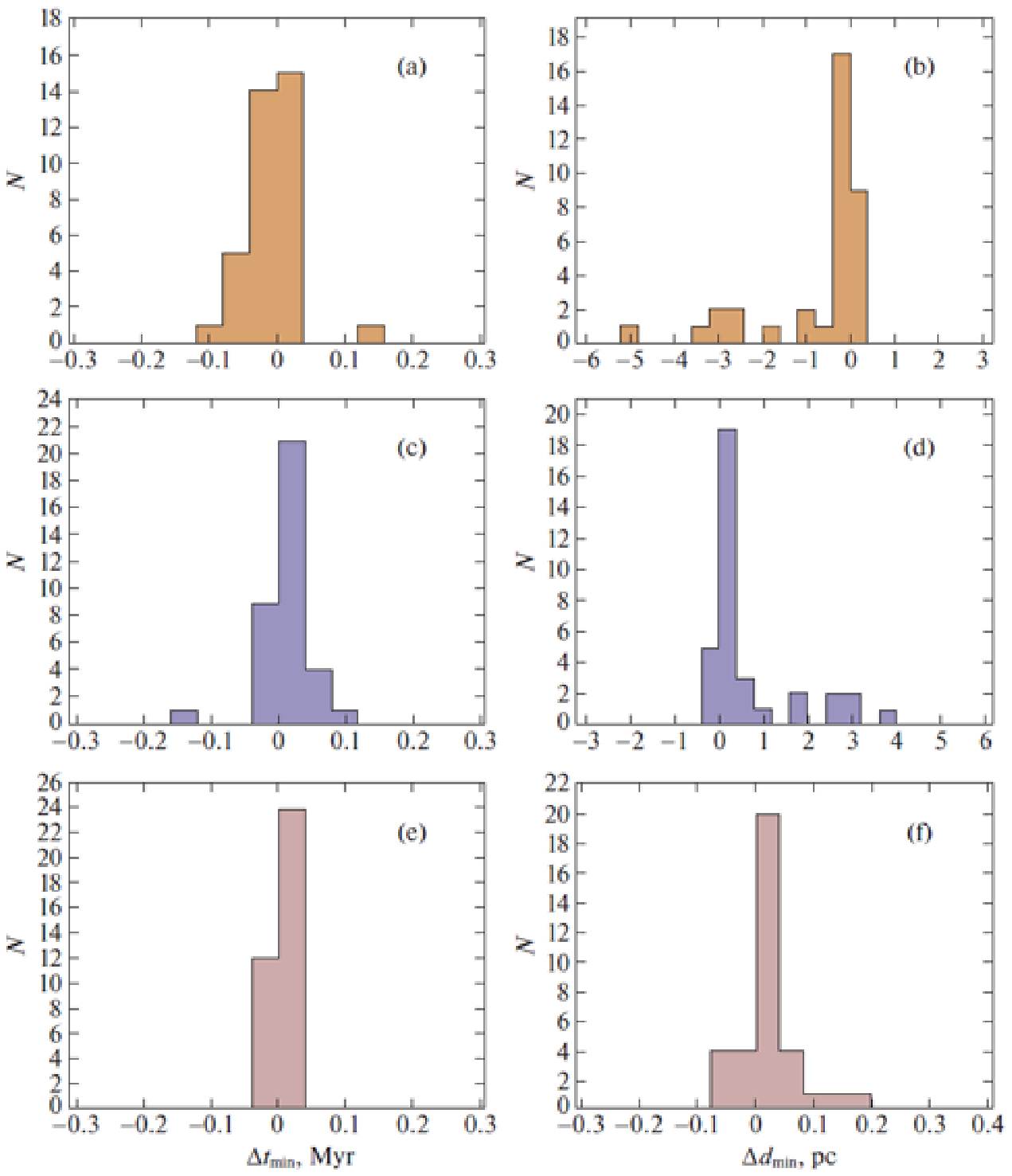}
 \caption{``Method 1 minus method 2'' (a, b), ``method 2 minus method 3'' (c, d), and ``method 1 minus method 3'' (e, f) encounter parameter differences. }
 \label{f-1} \end{center} }
 \end{figure}
%%%%%%%%%%%%%%%%%%%%%%%%%%%%
%%%%%%%%%%%%%%%%%%%%%%%%%%%%%%%%%%%%%%%%%%%%%%%%%%%%%%%%%% Fig.2
 \begin{figure} [t] {\begin{center}
 \includegraphics[width=130mm]{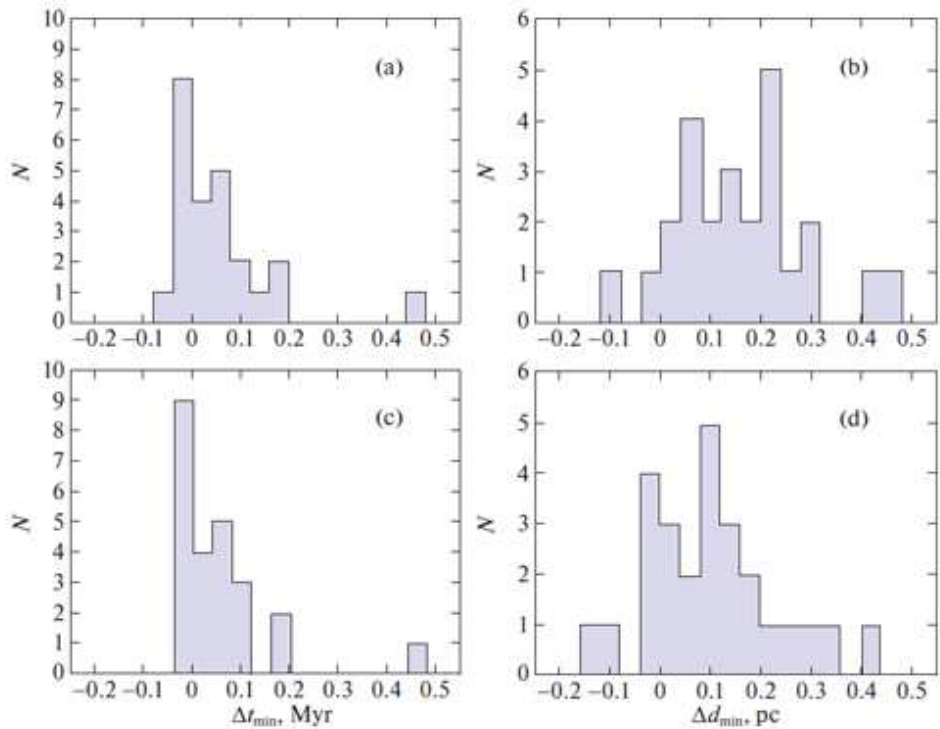}
 \caption{``Method 1 minus Bailer-Jones'' (a, b) and ``method 3 minus Bailer-Jones'' (c, d) encounter parameter differences. }
 \label{f-2} \end{center} }
 \end{figure}
%%%%%%%%%%%%%%%%%%%%%%%%%%%%

 \subsection*{RESULTS AND DISCUSSION}
Table 3 gives the parameters of the stellar encounters with the Solar system found by three methods:
(i) the linear one (2), (ii) the epicyclic one (3), and (iii) by integrating the orbits in the axisymmetric
potential (4). The last column gives the errors in the parameters (they can be attributed to all three
methods) estimated by the Monte Carlo method.

As can be seen from Table 3, there is excellent agreement between the encounter parameters found
by the first and third methods. In contrast, the estimates of the parameters obtained by the epicyclic
method occasionally have very strong deviations from those found by the two other methods. A correlation
of the deviations with the integration period is also easily seen: on a time interval longer than $\pm$1 Myr
the epicyclic method works poorly.

Based on the data from Table 3, we calculated the encounter parameter differences of the following
three types: ``method 1 minus method 2,'' ``method 2 minus method 3,'' and ``method 1 minus method 3.''
Figure 1 presents the histograms constructed from these differences.

On all three left panels (a), (c), and (e) the scale of the horizontal axis ($\Delta t_{\rm min}$) is the same. It can be easily seen that the ``method 1 minus method 3'' differences have the smallest dispersion. On panels (b) and (d) the scale of the horizontal axis ($\Delta d_{\rm min}$) exceeds the scale of panel (f) by an order of magnitude. Such long tails of the distributions on panels (b) and (d) arose due to the epicyclic method.

We may conclude that the encounter parameters found by methods 1 and 3 are in good agreement
between themselves. The strategy of searching for close encounters in which the linear method is applied
at the first, searching stage and the method of orbit integration in a potential is applied at the second,
more detailed stage is quite justified. For example, Bailer-Jones (2015) adhered to this strategy; the
model potential can be very complex and contain nonaxisymmetric components that take into account the
contributions of the spiral density wave or the central bulge (see, e.g., Garcia-S\'anchez et al. 2001).

For the two lower histograms in Fig. 1 ((e) and (f)) we obtained the following values of the mean and
its error: $\overline{\Delta t_{\rm min}}=-0.001\pm0.005$ Myr and $\overline{\Delta d_{\rm min}}= 0.037\pm0.125$ pc. The error of each of these methods will then be a factor of $\sqrt {2}$ smaller, $\overline{\sigma_{t_{\rm min}}}=0.003$ Myr and $\overline{\sigma_{d_{\rm min}}}=0.09$ pc. These values are smaller than the mean errors due to the contribution of measurement errors. For example, the following mean measurement errors were found from the data in the last two columns of Table 3: 
 $\overline{\sigma_t}= 0.085$ Myr and $\overline{\sigma_d}= 0.168$ pc.

Note that the encounter parameters found by us by the first and third methods are in agreement with
those from Bailer-Jones et al. (2018). We established this from 26 common stars. For this purpose, we
calculated the ``method 1 minus Bailer-Jones'' and ``method 3 minus Bailer-Jones'' parameter differences.
The results are given in Table 4. Our model potential differs from that of Bailer-Jones (2015) by
expression (7) for the halo. In addition, there are differing parameters for the coincident halo and disk
expressions (Table 1). Nevertheless, the ``method 3 minus Bailer-Jones'' differences, both $\Delta t_{\rm min}$ and $\Delta d_{\rm min}$, are small. Three or four stars, for example, Gaia DR2 3458393840965496960, represent an exception. There is also an example of a significant decrease in the differences $\Delta d_{\rm min}$ in the case of applying the third method compared to the first method, the star Gaia DR2 6608946489396474752.

The histograms constructed from the ``method 1 minus Bailer-Jones'' and ``method 3 minus Bailer-Jones'' encounter parameter differences are presented in Fig. 2. The following values of the mean and its
error were obtained for the ``method 1 minus Bailer-Jones'' differences: $\overline{\Delta t_{\rm min}}= 0.002\pm0.066$ Myr and $\overline{\Delta d_{\rm min}}= 0.162\pm0.141$ pc. The following values of
the mean and its error were obtained for the ``method 3 minus Bailer-Jones'' differences: $\overline{\Delta t_{\rm min}}= 0.069\pm0.070$ Myr and $\overline{\Delta d_{\rm min}}= 0.103\pm0.124$ pc. Here, three large ``outliers'' were discarded when estimating $\overline{\Delta t_{\rm min}}$ and its dispersion. On the whole, we may conclude that our method 3 is in slightly better agreement with the Bailer-Jones approach than with the linear one. More specifically:

(1) It can be seen from Table 3 that for the star ALS 9243 there is good agreement between the three
methods in determining the encounter time $t_{\rm min}\sim-2.3$~Myr. The epicyclic approach in estimating the
distance $d_{\rm min}$ is in poor agreement with the other methods. On the whole, we can confirm the conclusion
by Wysocza\'nska et al. (2020) that the star ALS 9243 is an interesting candidate for very close encounters with the Solar system. In fact, it is a candidate for a passage in the past through much of the Oort cloud.

(2) Based on the linear method,Dupuy et al. (2019) obtained the following estimates of the encounter parameters
for the star WISE J072003.20--084651.2AB:  $t_{\rm min}=-0.081\pm0.001$~Myr and $d_{\rm min}=0.333\pm0.010$~pc.
Our third method gives similar values,  $t_{\rm min}=-0.083\pm0.001$~Myr and $d_{\rm min}=0.370\pm0.012$~pc, while using the linear method, as can be seen from Table 3, we found parameters virtually coincident with those from Dupuy et al. (2009).

(3) Darma et al. (2019) obtained the estimates of $t_{\rm min}=-0.48\pm0.05$~Myr and $d_{\rm min}=0.58\pm0.11$~pc for the star Gaia DR2 52952724810126208 (designated there as ID 298) by integrating the orbits in an axisymmetric potential. According to our solution found by the third method, we have $t_{\rm min}=-0.553\pm0.060$~Myr and $d_{\rm min}=0.205\pm0.086$~pc. Thus, these results are in good agreement between themselves. The line-of-sight velocity of this star was taken from the Gaia DR2 catalogue, where  $T_{\rm eff}=6500^\circ~K$ and $\log g=3.5$~cm s$^{-2}$ are also given for this star, while its spectral type F0 is given in the LAMOST DR4 catalogue (Luo et al. 2018).

For the star Gaia DR2 3130033734235815424 (designated there as ID 291) Darma et al. (2019) found 
 $t_{\rm min}=0.49\pm0.05$~Myr and $d_{\rm min}=1.56\pm0.21$~pc. Our method in agreement with the
linear one, yields a completely different result, $t_{\rm min}=-0.620\pm0.116$~Myr and 
 $d_{\rm min}=0.299\pm0.096$~pc. Here, we see poor agreement with the results from Darma et al. (2019). Erroneous encounter parameters for this star are apparently given in Darma et al. (2019).

Note that its line-of-sight velocity was taken from the LAMOST DR4 catalogue (Luo et al. 2018).
Since the line-of-sight velocity has not been known previously, the star is of great interest for our problem.
The following parameters are also given there for this star: spectral type F0,$T_{\rm eff}=7150^\circ~K,$ and $\log g=3.948$~cm s$^{-2}$.

We may conclude that two stars of spectral type F0, Gaia DR2 52952724810126208 and Gaia DR2 3130033734235815424, are of great interest as candidates for close encounters with the Solar system. Each of them has a significant mass and, therefore, their passage through the Oort cloud could
produce noticeable perturbations of the comet cloud.

The encounter parameters for other common stars are not distinguished by good agreement. We trust
our results more, while the discrepancies are apparently related to the errors or misprints in Darma
et al. (2019).

 \subsection*{CONCLUSIONS}
In this paper we considered a sample of 36 candidates for close (within 1 pc) encounters with the
Solar system. The encounter parameters for this stars were calculated using the (1) linear and (2)
epicyclic methods and (3) by integrating the orbits in an axisymmetric potential. We concluded that
the epicyclic method works quite well only on a time interval no longer than $\pm$1 Myr, while the parameters
found by methods 1 and 3 are in excellent agreement between themselves. We concluded that in searching
for stellar encounters, the simple linear method could be applied at the first, searching stage and the more
complex method based on the integration of stellar orbits in a potential could be applied at the second,
more detailed stage.

We confirmed the conclusion by Wysocza\'nska et al. (2020) that the star ALS 9243 is an interesting
candidate for very close encounters with the Solar system. Based on the third method, we found the
following parameters:  $t_{\rm min}=-2.37\pm0.52$~Myr and $d_{\rm min}=0.21\pm0.80$~pc.

The encounter parameters for two stars from the list by Darma et al. (2019) are of interest. For
example, for the star Gaia DR2 52952724810126208 the following estimates were found by the third
method, in agreement with the linear one: $t_{\rm min}=-0.55\pm0.06$~Myr and $d_{\rm min}=0.21\pm0.09$~pc, while for the star Gaia DR2 3130033734235815424 $t_{\rm min}=-0.62\pm0.12$~Myr and $d_{\rm min}=0.30\pm0.10$~pc have been obtained for the first time.

On the whole, we showed that there are 15 candidates for encounters with the Solar system within 0.5 pc, i.e., candidates for a passage through the Oort cloud. The star GJ 710 still remains the record holder.
Based, for example, on the third method, we found the following estimates of the encounter parameters
for it: $t_{\rm min}=1.320\pm0.040$~Myr and $d_{\rm min}=0.016\pm0.009$~pc.

 \subsubsection*{ACKNOWLEDGMENTS}
We are grateful to the referee for the useful remarks that
contributed to an improvement of the paper.

 \subsubsection*{FUNDING}
This work was supported in part by Program KP19--270 of the Presidium of the Russian Academy
of Sciences ``Questions of the Origin and Evolution of the Universe with the Application of Methods of Ground-Based Observations and Space Research''.

 \bigskip \bigskip\medskip{\bf REFERENCES}{\small

1. E. Anderson and Ch. Francis, Astron. Lett. 38, 331 (2012).

2. C. A. L. Bailer-Jones, Astron. Astrophys. 575, 35 (2015).

3. C. A. L. Bailer-Jones, Astron. Astrophys. 609, 8 (2018).

4. C. A. L. Bailer-Jones, J. Rybizki, R. Andrae, and M. Fouesneau, Astron. Astrophys. 616, 37 (2018).

5. F. Berski and P. A. Dybczy\'nski, Astron. Astrophys. 595, L10 (2016).

6. V. V. Bobylev, Astron. Lett. 36, 220 (2010a).

7. V. V. Bobylev, Astron. Lett. 36, 816 (2010b).

8. V. V. Bobylev and A. T. Bajkova, Astron. Lett. 42, 1(2016a).

9. V. V. Bobylev and A. T. Bajkova, Astron. Lett. 42, 1 (2016b).

10. V. V. Bobylev and A. T. Bajkova, Astron. Lett. 43, 559 (2017).

11. V. V. Bobylev and A. T. Bajkova, Astron. Lett. 44, 676 (2018).

12. A. G. A. Brown, A. Vallenari, T. Prusti, de Bruijne, C. Babusiaux, C. A. L. Bailer-Jones, M. Biermann,
D. W. Evans, et al. (Gaia Collab.), Astron. Astrophys. 616, 1 (2018).

13. R. Darma, W. Hidayat, and M. I. Arifyanto, J. Phys.: Conf. Ser. 1245, 012028 (2019).

14. T. J. Dupuy, M. C. Liu, W. M. J. Best, A. W. Mann, M. A. Tucker, Z. Zhang, I. Baraffe, G. Chabrier, et al.,
Astron. J. 158, 174 (2019).

15. P. A. Dybczy\'nski, Astron. Astrophys. 396, 283 (2002).

16. P. A. Dybczy\'nski, Astron. Astrophys. 441, 783 (2005).

17. P. A. Dybczy\'nski and F. Berski, Mon. Not. R. Astron. Soc. 449, 2459 (2015).

18. F. Feng and C. A. L. Bailer-Jones, Mon. Not. R. Astron. Soc. 454, 3267 (2015).

19. B. Fuchs, D. Breitschwerdt, M. A. Avilez, C. Dettbarn, and C. Flynn, Mon. Not. R. Astron. Soc. 373,
993 (2006).

20. R. de la Fuente Marcos and C. de la Fuente Marcos, Res. Not. Am. Astron. Soc. 2, 30 (2018).

21. J. Garcia-S\'anchez, R. A. Preston, D. L. Jones, P. R. Weissman, J.-F. Lestrade, D. W. Latham, and
R. P. Stefanik, Astron. J. 117, 1042 (1999).

22. J. Garcia-S\'anchez, P. R. Weissman, R. A. Preston, D. L. Jones, J.-F. Lestrade, D. W. Latham, R. P. Stefanik, and J. M. Paredes, Astron. Astrophys. 379, 634 (2001).

23. J. Holmberg and C. Flinn, Mon. Not. R. Astron. Soc. 352, 440 (2004).

24. B. Lindblad, Ark. Mat., Astron., Fys. A 20 (17) (1927).

25. L. Lindegren, U. Lammers, U. Bastian, J. Hernandez, S. Klioner, D. Hobbs, A. Bombrun, D. Michalik, et al.
(Gaia Collab.), Astron. Astrophys. 595, A4 (2016).

26. L. Lindegren, J. Hernandez, A. Bombrun, S. Klioner, U. Bastian, M. Ramos-Lerate, A. de Torres, H. Steidelmuller, et al. (Gaia Collab.), Astron. Astrophys. 616, 2 (2018).
 
27. A.-L. Luo, Y.-H. Zhao, G. Zhao, et al., VizieR Online Data Catalog: V/153 (2018).

28. T. E. Lutz and D.H. Kelker, Publ. Astron. Soc. Pacif. 85, 573 (1973).

29. C. A. Martinez-Barbosa, L. J\'ylkov\'a, S. Portegies Zwart, and A. G. A. Brown, Mon. Not. R. Astron.
Soc. 464, 2290 (2017).

30. R. A. J. Matthews, R. Astron. Soc. Quart. J. 35, 1 (1994).

31. M. Miyamoto and R. Nagai, Publ. Astron. Soc. Jpn. 27, 533 (1975).

32. A. A. M\"ull\"ari and V. V. Orlov, Earth, Moon, Planets 72, 19 (1996).

33. J. F. Navarro, C. S. Frenk, and S. D. M. White, Astrophys. J. 490, 493 (1997).

34. J.H. Oort, Bull. Astron. Inst. Netherland 11 (408), 91 (1950).

35. T. Prusti, J.H. J. de Bruijne, A.G. A. Brown, A. Vallenari, C. Babusiaux, C. A. L. Bailer-Jones, U. Bastian, M. Biermann, et al. (Gaia Collab.), Astron. Astrophys. 595, 1 (2016).

36. I. A. Revina, Analysis of the Motion of Celestial Bodies and Estimation of the Accuracy of their
Observations (Latvian Univ., Riga, 1988), p. 121 [in Russian].

37. R. Sch\"onrich, J. Binney, and W. Dehnen, Mon. Not. R. Astron. Soc. 403, 1829 (2010).

38. The HIPPARCOS and Tycho Catalogues, ESA SP--1200 (1997).

39. S. Torres, M. X. Cai, A. G. A. Brown, and S. Portegies Zwart, Astron. Astrophys. 629, 139 (2019).

40. R. Wysocza\'nska, P. A. Dybczy\'nski, and M. Poli\'nska, arXiv: 2003.02069 (2020).
  }
  \end{document}